# Soft X-ray spectroscopies in liquids and at solid-liquid interface at BACH beamline at Elettra

S. Nappini[1], L. D'Amario[2,3], M. Favaro[4], S. Dal Zilio[1], F. Salvador[1], E. Betz-Güttner[5], A. Fondacaro[1], I. Píš[1], L. Romanzin[6], A. Gambitta[6], F. Bondino[1], M. Lazzarino[1], E. Magnano[1,7,a]

[1]IOM CNR laboratorio TASC, 34149 Basovizza (TS), Italy

[2]Freie Universität Berlin, Department of Physics Arnimallee 14, 14195 Berlin-Dahlem, Germany

[3]Uppsala University Dept. Chemisty Ångström, PO Box 523, 75120 Uppsala, Sweden

[4]Institute for Solar Fuels, Helmholtz-Zentrum Berlin für Materialien und Energie GmbH, Hahn-Meitner-Platz 1, 14109, Berlin, Germany

[5]Università degli Studi di Trieste, Physics Department, P.le Europa 1, 34127 Trieste, Italy

[6]Elettra-Sincrotrone Trieste, Area Science Park, 34149 Basovizza, Trieste, Italy

[7]Department of Physics, University of Johannesburg, PO Box 524, Auckland Park 2006, South Africa

Corresponding author: a)

*Dedicated to Sandi (Aleksander De Luisa, 1963-2019)*

**Abstract**

The Beamline for Advanced diCHroism (BACH) of the Istituto Officina dei Materiali-Consiglio Nazionale delle Ricerche (IOM-CNR), operating at Elettra synchrotron in Trieste (Italy), works in the extreme ultra violet (EUV)-soft X-ray photon energy range with selectable light polarization, high energy resolution, brilliance and time resolution. The beamline offers a multi-technique approach for the investigation of the electronic, chemical, structural, magnetic, and dynamical properties of materials. Recently one of the three end-stations has been dedicated to experiments based on electron transfer processes at the solid/liquid interfaces and during photocatalytic or electrochemical reactions. Suitable cells to perform soft X-ray spectroscopy in the presence of liquids and reagent gases at ambient pressure were developed. Here we present two types of static cells working in transmission or in fluorescence yield, and an electrochemical flow cell which allows to carry out cyclic voltammetry *in situ*, electrodeposition on a working electrode (WE) and to study chemical reactions *in-operando* conditions.

Examples of X-ray absorption spectroscopy (XAS) measurements performed in ambient conditions and during electrochemical experiments in liquid are presented.

**Introduction**

The global demand for energy and environmental sustainability requires the development of new methods to enable the comprehensive understanding of the electron transfer processes at the solid/liquid interfaces. Photocatalytic and electrochemical reactions, such as $CO_2$ reduction reaction ($CO_2$RR), oxygen reduction reaction (ORR), oxygen evolution reaction (OER), and hydrogen evolution reaction (HER) represent a milestone in the current concepts of sustainable energy research.[1,2] A rational design of new catalysts and (photo) electro catalysts must rely on detailed understanding of the reaction mechanisms occurring at the surface of the catalysts. Spectroscopic techniques based on synchrotron radiation are pivotal to provide significant insights into the behaviour of electrocatalysts under operating conditions. In particular, an exhaustive comprehension of the electronic properties of materials in ambient pressure gas or in liquid environment is required, as well as mechanistic understanding of the electronic evolution occurring at the solid/liquid or solid/gas interfaces during catalytic or electrochemical reactions.

In order to characterize such interfaces *in situ*, one would need a probe that can penetrate the solid and/or the liquid layer while generating collectable signals from the interface layer (typically several nanometers thick) or from the bulk region of the solution. Therefore, there is an increasing demand to develop techniques or to implement the existing ones to study phenomena at the interfaces under *in situ* or *operando* conditions.

Several *in situ* cells have been recently designed for various synchrotron-based X-ray techniques, such as X-ray diffraction (XRD)[3,4] scanning/transmission X-ray microscopy (STXM/TXM)[5,6], X-ray spectroscopies (absorption and photoemission) in the hard X-ray range[7,8] and in the soft X-ray energy range.[9,10,11,12]

Among the X-ray spectroscopic techniques, EUV and soft X-ray spectroscopies(SXS) are characterized by intrinsic surface sensitivity which allows studying physical and chemical processes occurring at the material surfaces and interfaces. Moreover, SXS techniques enable to investigate the L-edges of the first-row transition metals, $M_{5,4}$ edges of the rare-earth elements, and K-edges of light elements such as Li, Be, B, C, O, N, and F. The absorption spectroscopy is element and chemical-state sensitive and provides information about the unoccupied electronic states with few to hundreds of nanometer probing depth, allowing the study of variations in the electronic structure of the materials, such as the oxidation state and ligand field strength, which change during the electrochemical and energy conversion processes involved in fuel cells or batteries.

The electronic structure can be probed by photon-in/electron-out techniques like X-ray photoelectron spectroscopy (XPS), or X-ray absorption (XAS) in total electron yield mode (TEY), and by photon-in/photon-out techniques such as XAS in total fluorescence yield mode (TFY), and X-ray emission spectroscopy (XES).

These techniques typically require ultra-high vacuum (UHV) conditions, making difficult their application to the study of catalytic and (photo)electrochemical processes under operating environments. For this reason, it is important to bridge the "pressure gap" between UHV and ambient pressure by integrating SXS techniques into ambient or near-ambient pressure environments.

So far, different approaches have been developed.[13] The first is near ambient-pressure photoelectron spectroscopy (NAP-XPS)[14,15,16,17,18,19] in which the sample is kept under conditions close to the ambient environment, and in some cases a very thin liquid layer is condensed on top of the sample. This setup requires a differentially pumped electrostatic pre-lens system, with a three-stage differentially pumped analyzer. The second approach is the use of liquid microjets/droplet trains in combination with differentially pumped photoelectron analyzers and high brilliance synchrotron radiation sources for *in situ* XAS and XPS measurements.[20] This system allows the solid/liquid interface characterization at the nanoparticle surface, but it is not suitable for electrochemistry experiments. The main advantages in using liquid microjets or droplet trains are the negligible beam damage and sample contamination. In addition, the use of a liquid droplet train in combination with a delay-line detector enables performing time-resolved studies, with the possibility to explore time scales spanning from seconds down to microsecond.

The third approach is based on the confinement of a liquid or a gas in environmental cell (E-cells) with windows separating the fluid from the vacuum which exploit the ability of photons to travel deeper in solid materials (up to thousands of Å) . The window material has to be chosen appropriately as it should be inert toward the cell components, such as electrodes and electrolytes, it should be resistant to the oxidation and reduction during charge processes, be mechanical resistant to withstand large pressure differences, and be transparent enough to the X-rays. For these reasons a thin $Si_3N_4$ membrane is usually used, but recently also graphene membranes have been successfully employed to confine the liquid in the cell[21,22] or on an appropriate substrate.[23,24]

Several types of cells have been developed: static[25] or microfluidic,[26] working at high liquid pressure[11] in transmission[27,9] or in reflection geometry.[28]

Following the last approach, two types of cells dedicated to soft X-ray absorption spectroscopy measurements have been developed to fit the BACH beamline parameters: 1) static cells working in transmission and in fluorescence yield; 2) microfluidic cells dedicated to *operando* electrochemical experiments. The cells were developed and fabricated in house by the technical group of the IOM-CNR. In the present work, we report a detailed description of the developed cells together with some proof-of-concept measurements performed at the O K-edge in various liquid solutions and transition metal (TM) L-edge spectra obtained during electrochemical experiments in the microfluidic cell.

# 1. The in situ/operando cells

## 1.1 Static liquid cells (transmission and fluorescence yield)

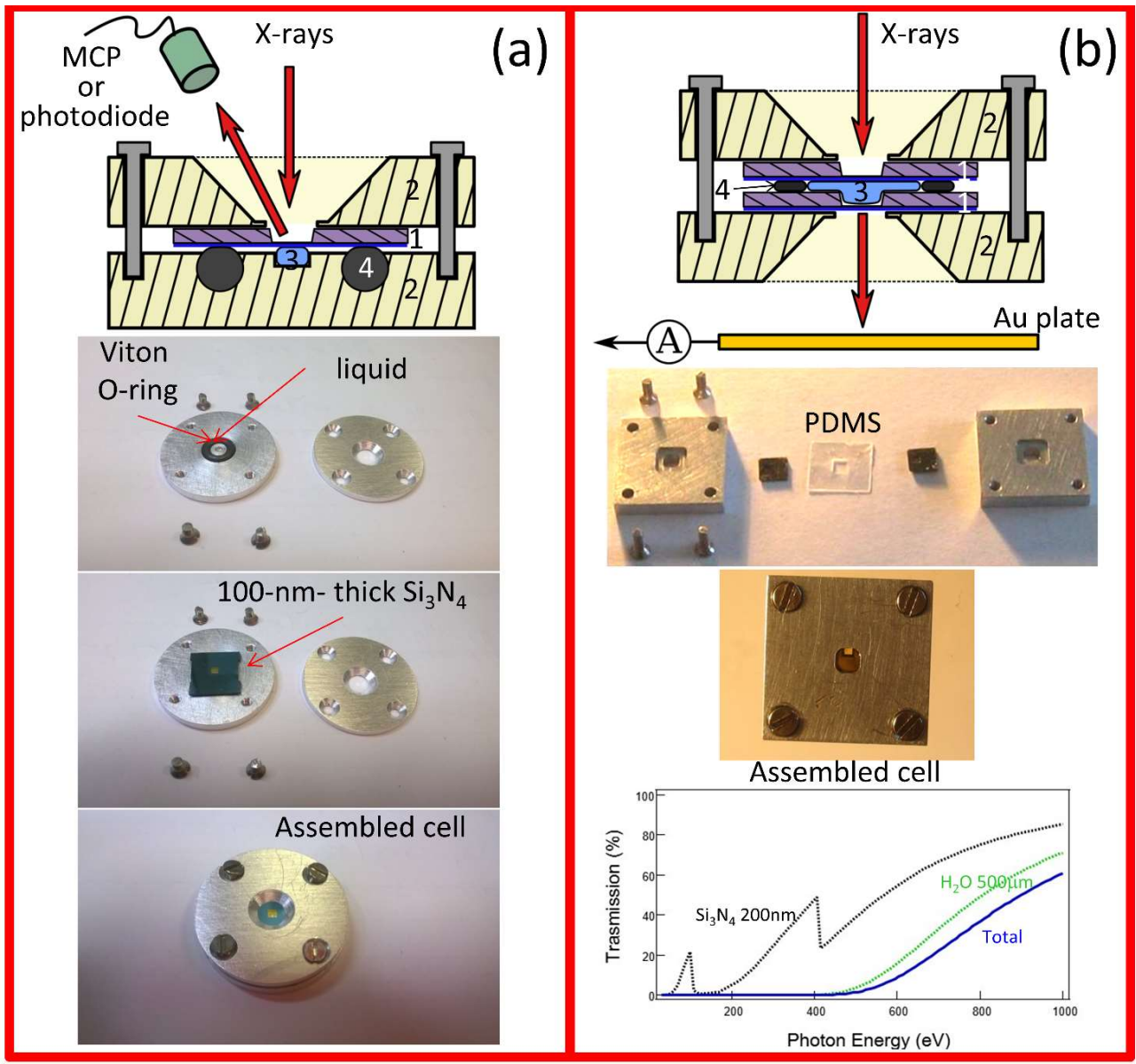

Fig. 1: (a) Static liquid cell for XAS in fluorescence yield. On top a scheme of the cell and the used experimental geometry is reported. A 100 nm-thick-$Si_3N_4$ window (1) pressed between two aluminum or PEEK plates (2) closes the liquid in a small cavity (3). A Viton O-ring (4) ensures the sealing of the cell. The pictures show the components of a cell made of aluminum. (b) Static liquid cell for XAS in transmission. On top a scheme of the cell and the used experimental geometry. Two 100 nm-thick-$Si_3N_4$ windows (1) pressed between two aluminum plates (2) encapsulate a small volume of liquid of ~ 500μm thickness (3). The seal is guaranteed by a 100μm-thick PDMS layer (4). The signal is collected by a picoammeter connected to an Au plate. The pictures show the components of the cell. On bottom the total transmission curve (blue line), calculated considering 200nm-thick $Si_3N_4$ (black line) and 500 μm-thick liquid $H_2O$ (green dotted line), is shown.

Two types of static liquid cells (presented in Fig. 1) have been developed with the aim of using two photon-in/photon-out detection modes. The first cell ((a)- cell 1) works in fluorescence yield and it consists of a thin X-ray transparent membrane which separates the vacuum chamber from the liquid reservoir. The second type of cell ((b)- cell 2) operates in transmission mode and it consists of two thin membranes clamped together to form an inner space where the liquid is confined.

The used membranes are typically made of $Si_3N_4$, but different materials, such as $SiO_2$ or SiC, could be used to optimize the transmission for the absorption edges under investigation. The typical membrane window dimensions are 1×1 mm$^2$ with a thickness of 100 nm, and the silicon frame is 1x1 cm$^2$ wide with a thickness of 200-500 μm. These parameters were chosen as a compromise between the X-ray transmission and the mechanical strength of the membrane itself.

The calculated transmission curve[29] considering $Si_3N_4$ of a total thickness of 200 nm is reported in figure 1 (b) (bottom panel, black dashed line). The curve reproduces the signal transmitted through two 100-nm-thick windows, excluding the presence of any medium inside the cell. In the soft X-ray energy range of interest (> 250 eV to probe also C K-edge at ~285–290 eV) the transmission is higher than 20% which permits to achieve a good signal-to-noise ratio in the acquired spectra in a reasonable time.

The $Si_3N_4$ membranes were fabricated at the Nanofabrication IOM-CNR facility following standard procedures of photolithography.[30]

In details, for the cell 1 (fig. 1 (a)): the $Si_3N_4$ membrane (1) separates the liquid volume of around 3-5 μl (3) from the UHV experimental chamber. The Viton O-ring (4) guarantees the appropriate seal to work in UHV condition (base pressure <1x10$^{-8}$ mbar). The body of the cell is made of aluminum or polyether-ether ketone (PEEK), an UHV compatible chemically inert polymeric material, with good mechanical properties and electrical insulation. The cell is mounted on the standard sample carrier (VG XL25VH or XL25HC) available on the end-station "A" of BACH beamline, inserted in a load lock and then transferred to the measurement chamber.

The cell in the load lock is pumped down to a pressure $<1\times10^{-6}$ mbar, first by a roughing pump at a pumping rate around 50 mbar/min from atmospheric pressure down to ~1 mbar and then to high vacuum with a turbo pump. Afterwards it is transferred to the measurement chamber (base pressure $10^{-9} - 10^{-10}$ mbar). The load lock allows to reach the operating conditions in a relatively short time (usually 1 h) and it ensures that the pressure in the experimental chamber during the measurement is lower than $5\times10^{-8}$ mbar. In a similar way, the cell can be taken off easily paying attention to the $N_2$ venting operation, which must be done slowly from ~1mbar to atmospheric pressure (rate ~ 50 mbar/min).

In fig. 2 we report O K-edge total fluorescence yield XAS spectra obtained from various samples in environmental conditions enclosed in the static cell. The measurements were performed using radiation linearly polarized in the horizontal plane, an angle of incidence of the X-rays normal to the cell window and the spectra were acquired in fluorescence yield by an MCP (Hamamatsu F4655-13) detector at an angle of 22.5° degrees from the normal to the membrane. The size of the beam is 300 μm (horizontally) x 100 μm (vertically).

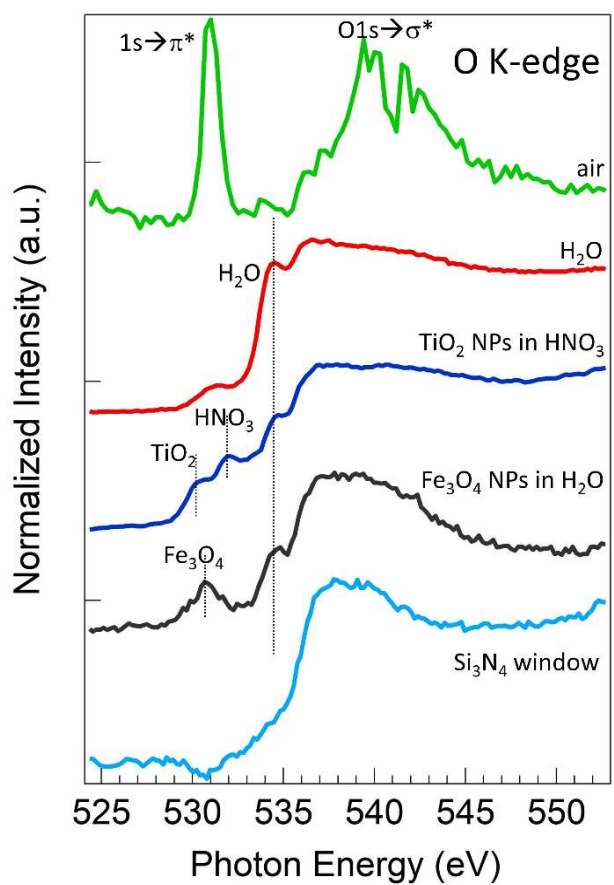

Fig. 2: XAS O K-edge of $Fe_3O_4$ NPs in $H_2O$ (black line), $TiO_2$ NPs in $HNO_3$ (blue line), $H_2O$ liquid (red line), air (green line) encased in the static cell measured through a100nm-thick $Si_3N_4$ window in total fluorescence yield. The O K-edge signal from the $Si_3N_4$ window (light blue line) is reported as a reference.

Figure 2 shows the O K-edge spectra from $Fe_3O_4$ nanoparticles (NPs) dispersed in water (black line), $TiO_2$ NPs in an aqueous solution of $HNO_3$ (blue line), liquid $H_2O$ (red line), and air (green line). As a reference for the background, the O K-edge signal from the $Si_3N_4$ window is also reported (cyan line). The spectra of $Fe_3O_4$ NPs, $TiO_2$ NPs, and $H_2O$ show the characteristic pre-edge of liquid water at 535 eV photon energy associated with broken or weakened H-bonds, the main edge around 537 eV assigned to the unsaturated hydrogen bonds, and a post edge around 537–545 eV associated with the saturated hydrogen bonding network in the bulk regions of the liquids.[28,31] Additional features at 530 eV can be ascribed to 3d- metal orbitals hybridized with O-2p orbitals in the crystal lattice of the NPs ($TiO_2$ and $Fe_3O_4$) or to unwanted contamination of $Si_3N_4$ window in the $H_2O$ spectrum. The peak at 532 eV of $TiO_2$ NPs in $HNO_3$ solution corresponds to interfacial $NO_3$-species.[32] For comparison, theO K-edge XAS spectrum obtained from the air closed in the static cell is shown (top spectrum). It is characterized by a strong resonance at 531.2 eV (O1s→π* transition) and a broad absorption band in the energy range 537-545 eV (O1s→σ* transition) typical of $O_2$ gas.[33]

For the cell 2 (see fig. 1 (b)) two $Si_3N_4$ membranes (1) are sealed together by a thin layer of polydimethylsiloxane (PDMS) (4) and the enclosed liquid volume is around 0.5 µl (3), with about 500-µm-thick liquid region. The cell (mounted on a 3-degree of freedom manipulator) is operated in the measurement chamber of the central branch of the beamline BACH. The XAS measurements are performed in transmission geometry by collecting the drain current on a gold foil placed on the back of the cell at a distance of few cm. The gold foil is connected to a picoammeter, as sketched in fig. 1 (b) (top panel), via a BNC feedthrough. The pumping/venting operations used to load the cell 2 are similar to those described for the cell 1. In this case the turbo pump allows to reach a vacuum $<1\times10^{-7}$ mbar in 5-6 hours. A liquid $N_2$ trap is also used to lower the pressure in the chamber and to reach the operating pressure $<5\times10^{-8}$ mbar.

The calculated transmission curve[29] for this cell is reported in figure 1 (b) (bottom panel, solid blue line), considering the transmission of the two $Si_3N_4$ windows of 100 nm thickness each, and the transmission through a 500 µm thick $H_2O$ layer. The cell was tested encapsulating an aqueous solution of $FeCl_3$ 40 mM and the transmission signal, measured in the 450–750 eV photon energy range, is reported in Fig. S1 (red circle points), showing a good agreement between the experimental data and the calculated total transmission signal.

## 1.2 Microfluidic electrochemical cell

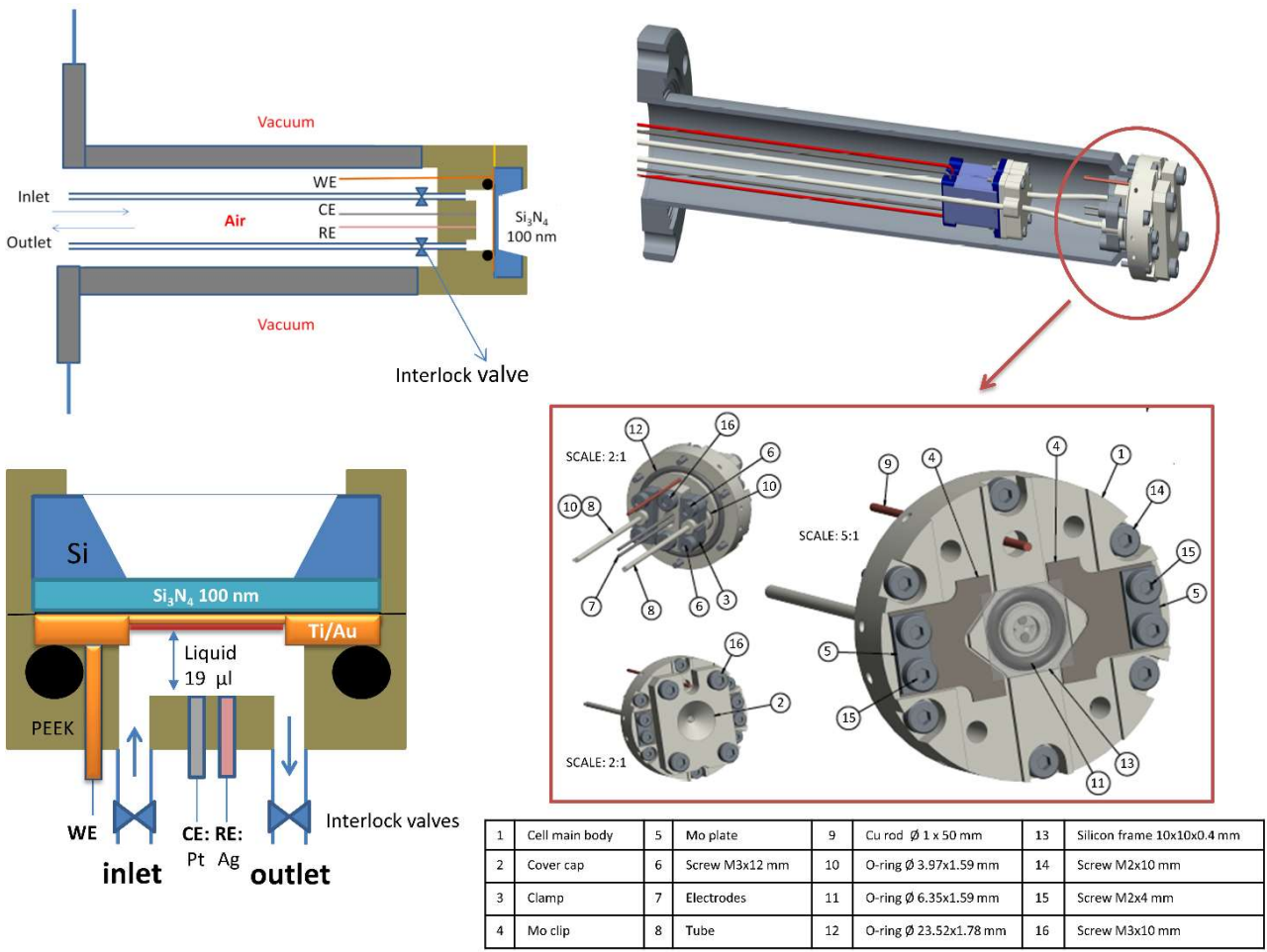

Fig. 3: Schematic drawings of the microfluidic cell for *in-operando* XAS measurements. The cell is equipped with a three electrodes system where the $Si_3N_4$ membrane is the working electrode, a 1mm-diameter Pt wire is the counter electrode, and a 1mm-diameter Ag wire is the quasi-reference electrode. The body of the cell in PEEK is sealed by a Viton O-ring on a tubeinserted into the UHV chamber. The interlock valves are mounted on the microfluidic channels and allow to limit the volume of the liquid in the cell during measurements.

One of the main limitationsof the static cells is the possible degradation of the liquid and the formation of radicals during X-rays irradiation. Furthermore, the electrolyte can be changed only by taking off the cell and breaking the vacuum of the chamber. A new cell equipped with microfluidic channels, which can provide fresh solution and guarantee mixing and good diffusion of the liquid, was developed to overcome the problems mentioned above. Moreover, three electrical contacts were added in order to perform *operando* XAS measurements during electrochemical experiments.

A schematic drawing of the cell is shown in figure 3. More technical details of the cell are reported in Fig. S2.

Similarly to the static cell, a thin $Si_3N_4$ membrane separates the liquid in the cell (made in PEEK) from the vacuum of the experimental chamber. The liquid is supplied through two PEEK tubes with an inner diameter of 250 μm, one for the inlet and one for the outlet. Both tubes are supplied with 24 V whisper microvalves (Bürkert) interlocked with the chamber pressure gauge. In case of a pressure increase above $1\times10^{-6}$ mbar in the experimental chamber (for example in case of accidental windows breaking) the valves close the liquid flux, thus limiting the liquid volume leaking into the vacuum chamber to only 19 μl, i.e. the total volume of the cell (which includes the tubes between the cell body and the microvalves).

The liquid is inserted in the tube through a syringe of 1 ml volume moved by a pump (NE-1000 Programmable Single Syringe Pump) which allows a controlled liquid flow rate in the range between $2\cdot10^{-4}$ and 590 μl/s. A typical flow rate used in our experiments was 0.83 μl/s, which corresponds to a complete replacement of the liquid in 155 seconds. The flow prevents the liquid samples from the radiation damage during measurements and allows to perform experiments which require the use of different solutions without breaking the vacuum conditions.

The electrical contacts are used to carry out electrochemical experiments such as cyclic or linear sweep voltammetry (CV, LSV), electrochemical depositions and electrochemical impedance spectroscopy (EIS) while simultaneously acquiring the XAS spectra *in operando* fashion.

The working electrode (WE) is the $Si_3N_4$ membrane itself coated with a Ti/Au (3nm/15nm) layer (thereby assuring good electrical conductivity). The membrane is pressed by a PEEK top cap on two molybdenum plates which are electrically connected to the air side through a Cu wire. Electric conductive and chemical resistant molybdenum clamps have been specifically designed to ensure a homogenous current flow on the entire $Si_3N_4$/Ti/Au membrane. The quasi-reference electrode (Ag) and the counter electrode (Pt) are two wires of 1 mm diameter placed on the bottom of the cell in contact with the solution in the cell reservoir. The electrochemical measurements were performed using an Emstat3+ (PalmSens) potentiostat.

In order to test the performance of the new microfluidic cell under standard conditions and to find the best operating parameters, a 3D printed cell (printer: Formlabs, Form2, SilkyGenie) (figure 4(a))was fabricated using a clear resin (FLGPCL04) with good chemical stability. The printing resolution was around 100 μm. The cell was redesigned starting from the previous one, adapting it for the use in air. An exploded 3D drawing is reported in fig. 4(b).For more details of the cell design see figure S3.The 3D printed cell was successfully tested for several applications, such as electrodeposition of catalytic materials on the Au/Ti coated $Si_3N_4$ window (WE) and cyclic voltammetry measurements. As example, the measured current and voltage during a Ni hydroxide $(Ni(OH)_2)$ cathodic electrodeposition is reported in fig. 4(c). The experiment was performed following a similar procedure as described in the next paragraph.

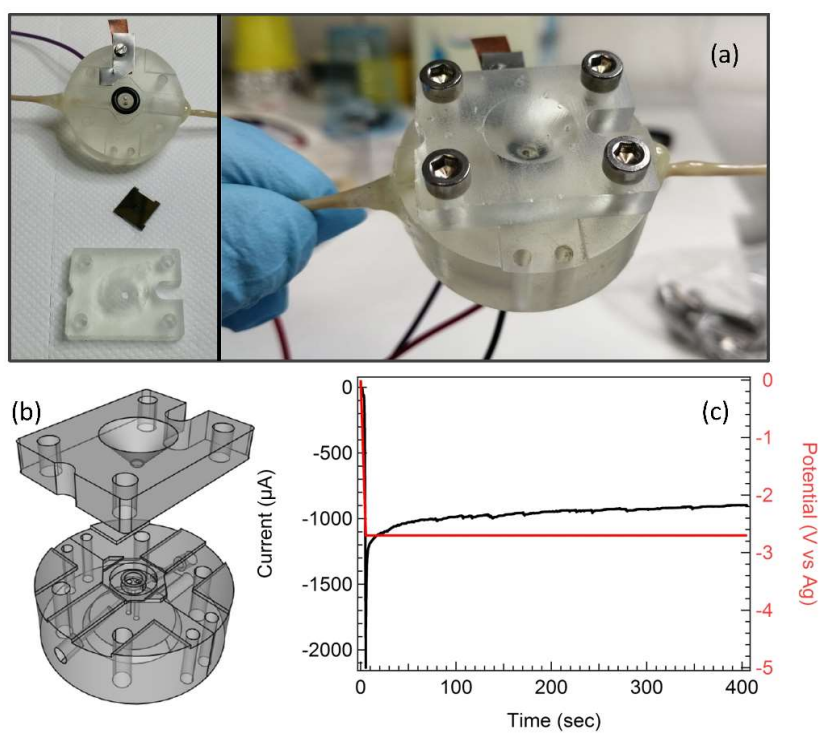

Fig. 4: (a) Pictures of the 3D printed cell used to test the cathodic electrodeposition of catalytic materials on $Si_3N_4$ and to carry out preliminary cyclic voltammetry measurements to optimize the operating conditions required for *in-operando* experiments. (b) 3D exploded drawing of the cell. (c) Plots of the current (black line) and the applied Voltage (red line) vs time during Ni hydroxide $(Ni(OH)_2)$ cathodic electrodeposition.

## 2. Electrochemical experiments

### 2.1 Electrochemical deposition of Ni(OH)$_2$ and cyclic voltammetry

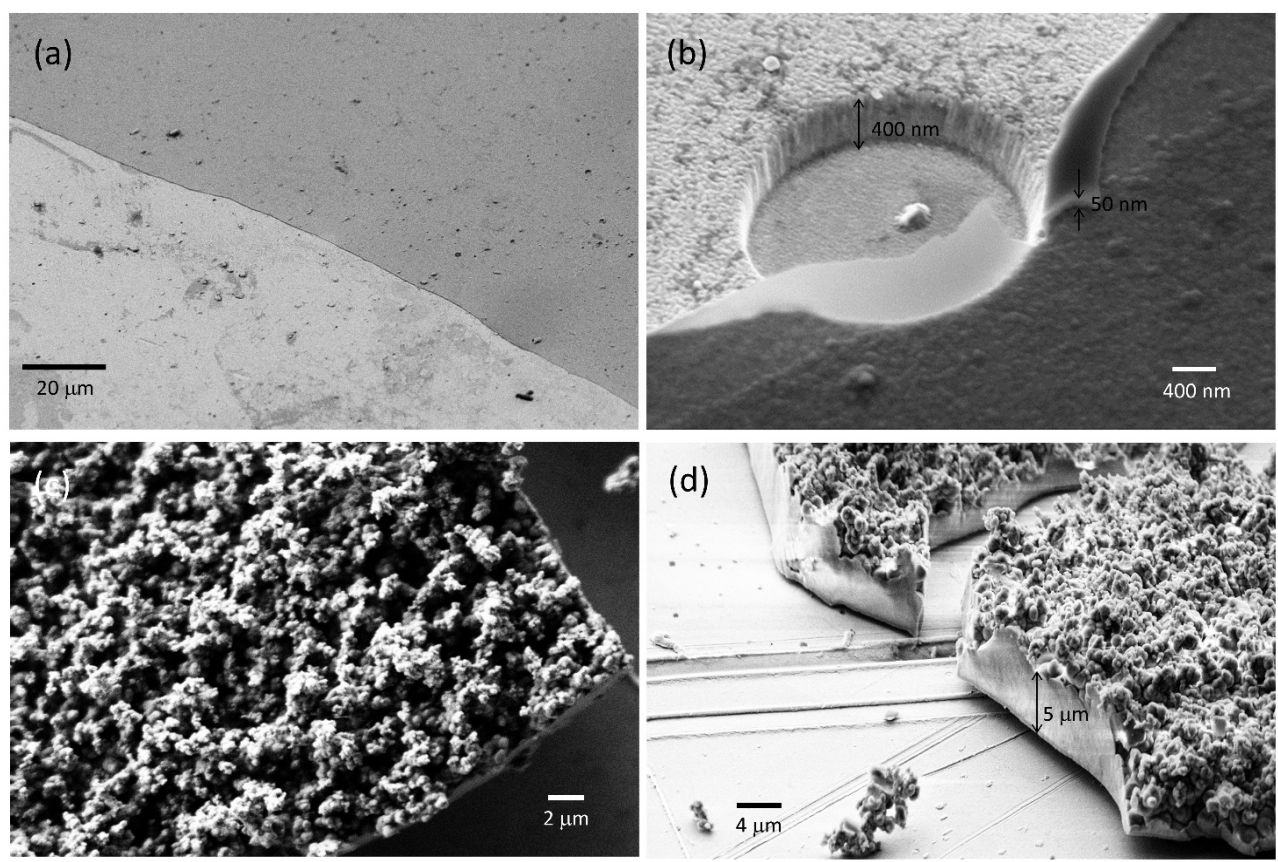

Fig.5: (a) SEM images of Ni(OH)$_2$ layers obtained by cathodic electrodeposition and subsequent activation on Si$_3$N$_4$ 100 nm-thick membrane coated with a Ti(3nm)/Au(15 nm) (see text for experimental details). (a) and (b): SEM images of a ~ 400 nm-thick film; (c) and (d): SEM images of a ~5 μm-thick film.

The electrochemical microfluidic cell was successfully used for cathodic electrodeposition of Ni(OH)$_2$, which can be electrochemically converted into the Ni catalytic-active phase composed by Ni oxide-hydroxide (NiOOH)[34] in alkaline medium, one of the most promising catalysts for OER, especially if combined with Fe-based materials, which are able to enhance its electro-catalytic activity.[35,36,37] Two Ni(OH)$_2$ films with different thickness were cathodically electrodeposited on the Si$_3$N$_4$ (100 nm) membrane coated with a Ti(3nm)/Au(15 nm) conductive layer (working electrode)[38] from an aqueous solution containing 0.1 M Ni(NO$_3$)$_2$× 6 H$_2$O and 25 mM (NH$_4$)$_2$SO$_4$ at pH 2.5. In fig. 5 we report the SEM images of two samples deposited by applying an average cathodic current of 0.9 mA/cm$^2$ for 60 s (thin layer, images (a) and (b)) and 2.3 mA/cm$^2$ for 200 s (thick layer, images (c) and (d)) on a circular area of 0.31 mm$^2$. The thickness was estimated to be 400 nm (thin) and 4μm (thick) assuming 0.125 mC/cm$^2$ per 1 nm [39]. The images were acquired using a SEM Zeiss supra40 instrument, with 10 nm lateral resolution, and voltage set to 2–5 kV. The thin film appears uniform (fig. 5(a)), composed of a dense layer of pillars around 400 nm high, partially covered by a second thinner layer (50 nm), as evidenced by the circular defect reported in the high resolution image (fig. 5(b)) with the aim to show clearly the morphology and the thickness of the sample). The thick layer reveals a double structure: a very dense one at the bottom and a second porous 'cauliflower' like structure on the top, with a total thickness around 5 μm(fig. 5(c) and (d)).

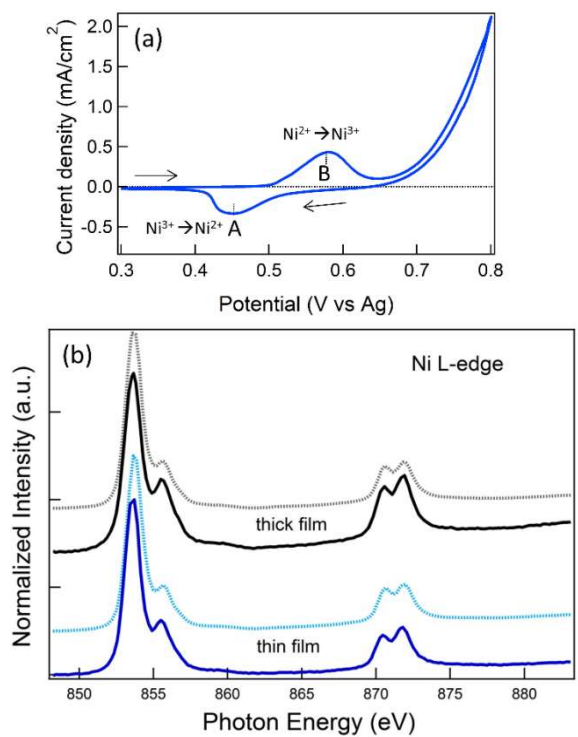

Fig. 6: a) Example of cyclic voltammetry measured in 0.1 M KOH on an electrodeposited Ni(OH)$_2$ layer (WE) vs an Ag wire pseudo-reference electrode; b) XAS Ni L-edges measured in total electron yield (dash lines) and total fluorescence yield (solid lines) on the Ni(OH)$_2$ thin (bottom curves) and thick layers (top curves).

In Figure 6 (a) a CV curve (recorded in the 0.3 – 0.8 V vs. Ag potential range) of the Ni(OH)$_2$/NiOOH coated electrode in 0.1 M KOH is reported. The CVs collected in this work are comparable to those found in literature for similar materials[35,36,40,41].

The voltammogram in Figure 6(a) displays two red-ox peaks centered around 0.45 V (A) and 0.58 V (B) in the cathodic and anodic traces, respectively. The redox waves are likely to be attributed to the following reactions:

I) anodic reaction 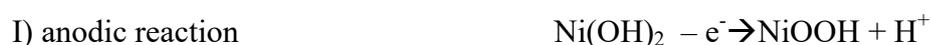 $Ni(OH)_2 - e^- \rightarrow NiOOH + H^+$

II) cathodic reaction 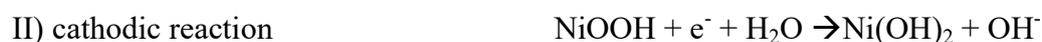 $NiOOH + e^- + H_2O \rightarrow Ni(OH)_2 + OH^-$

Thus, the peaks A and B can be attributed to the oxidation of $Ni^{2+}$ to $Ni^{3+}$ in the anodic direction and the corresponding reduction in the cathodic scan.

The Ni(OH)$_2$ catalyst was characterized *ex-situ* by XAS, measuring the Ni L-edges on a separate UHV end-station of BACH beamline. The spectra (see figure 6 (b)) were measured both in surface sensitive total electron yield (dash lines) and in more bulk sensitive total fluorescence yield mode (solid lines). The photon energy was calibrated using XPS (available in the same chamber) and measuring Au 4f on a reference sample in electrical contact with the sample. Ni L- line shape results from the 2p→ 3d dipole transitions, split in two regions for the core-hole spin-orbital interaction, the L$_3$ around 853 eV (2p$_{3/2}$→3d) and the L$_2$(2p$_{1/2}$→3d) around 870 eV. The features in both L$_3$ and L$_2$ edges originate from atomic multiplet effects and are sensitive both to the electronic and oxidation state of the metal, and to the local geometry of Ni atoms. In this case Ni L-edge shows the typical features of a 2+ oxidation state.[42]

## 2.2 *Operando* XAS measurements on Fe-Ni catalyst

Fig. 7 (a) shows the set-up for *operando* experiments. The microfluidic cell has been designed to fit the end-station of the central branch line of the BACH beamline. The cell tube is mounted horizontally (parallel to the ground plane)in the experimental chamber on a port-aligner which allows the optimization of the position of the cell horizontally and vertically (perpendicular to the ground plane) (see picture fig. 7 (b)). TheX-ray beam is focused on the cell by a toroidal mirror (TM). The TM is mounted tilted by 2.5°, with respect to the incoming beam, facing down, with a deflecting angle of 5°. After the TM, the light is tilted by 3° with respect to the horizontal plane, pointing downward. The use of only one reflecting element limits the loss of flux, caused by the reduced reflectivity of the optical elements. Two detectors were used for the XAS measurements in total fluorescence yield: 1) an Hamamatsu MCP, with a background signal of a few counts/s, working at -1600 V supply voltage and $10^6$ gain; 2) an AXUV 100 IRC Photodiode. The two detectors are placed symmetrically in front of the cell at an angle of around 20° with respect to the incident direction of the x-ray beam. The incoming flux was monitored by measuring the photocurrent from an Au grid mounted at the beginning of the beamline end-station.

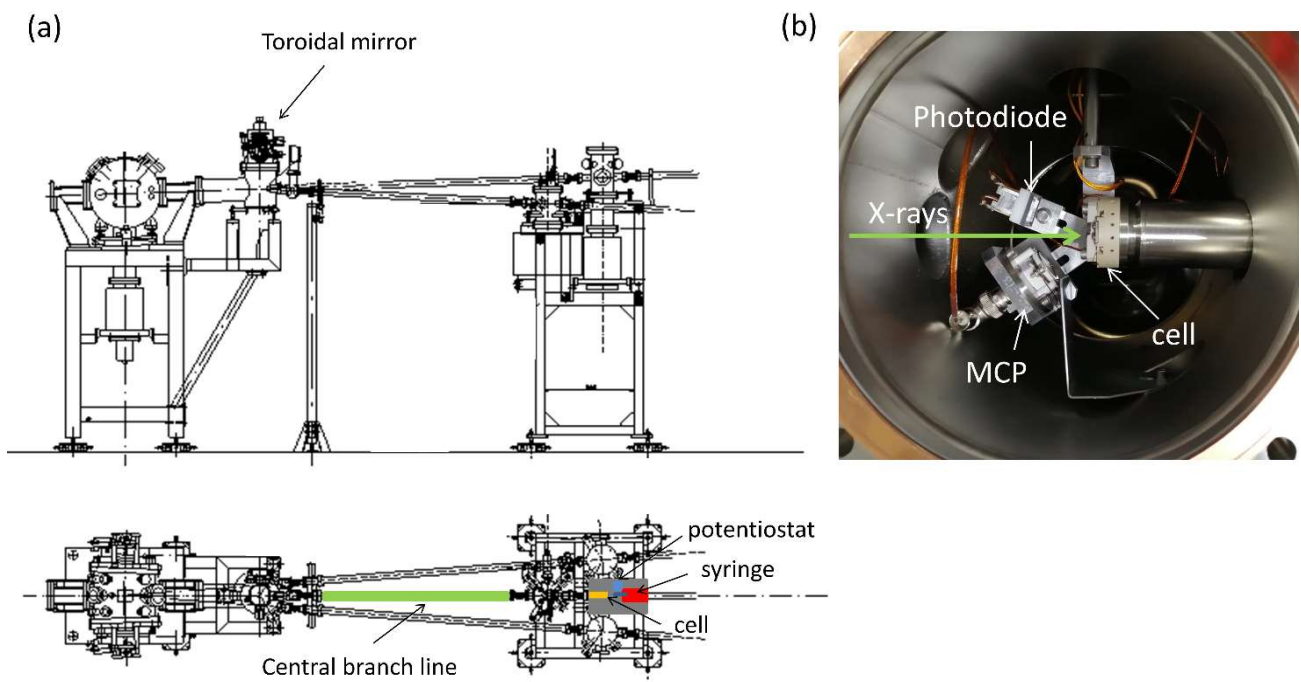

Fig. 7: Set-up for *in operando* experiments; a) side and top schematic view of the toroidal mirror chamber and the endstation; b) picture of the experimental set-up where the Photodiode and MCP detectors are mounted at around 20° from the incident beam direction at around 4 cm from the cell.

As an example of *operando* experiments, we followed, by XAS measurements on the Ni L-edge, the redox processes of a Fe-Ni catalystpreviously deposited on the $Si_3N_4$/Ti/Au membrane. The sample was cathodically electrodeposited as $Fe_xNi(OH)_2$ following a similar procedure to the one described in the previous paragraph, using the same solution with the addition of $FeSO_4$ to get the desired percentage of Fe (10 mol%). To define the area of deposition and increase the thickness homogeneity throughout the sample, the $Si_3N_4$/Ti/Au membrane was masked by an inert plastic mask (see figure S4)while immersed in the electrolyte deposition solution. A fixed cathodic current of 20mA/cm$^2$was applied for 5 seconds using a Biologic SP-240potentiostat/galvanostat. The deposited layer thickness was estimated to be 400 nm.

$Fe_xNiOOH$, formed during the first CV cycle, is known to show a good volume activity, which is the property attributed to solid state catalysts that are able to perform catalytic reactions throughout their entire solid volume.[43,44,45]The volume activity of the catalyst is very important for *in situ* soft

X-rays XAS measurements because of the limited probing depth at this energy range. In fact, the active catalyst surface must be reached by the x-rays through the window thickness and the film layer, and then the emitted photons must be transmitted back to the detector.

Ni L-edge XAS spectra were measured at open circuit potential in total fluorescence yield using the two detectors (MCP and photodiode) (see figure S5). They reveal a typical shape of a mostly Ni $^{2+}$ oxidation state[42], as confirmed by Ligand Field Multiplet simulations carried out by CTM4XAS

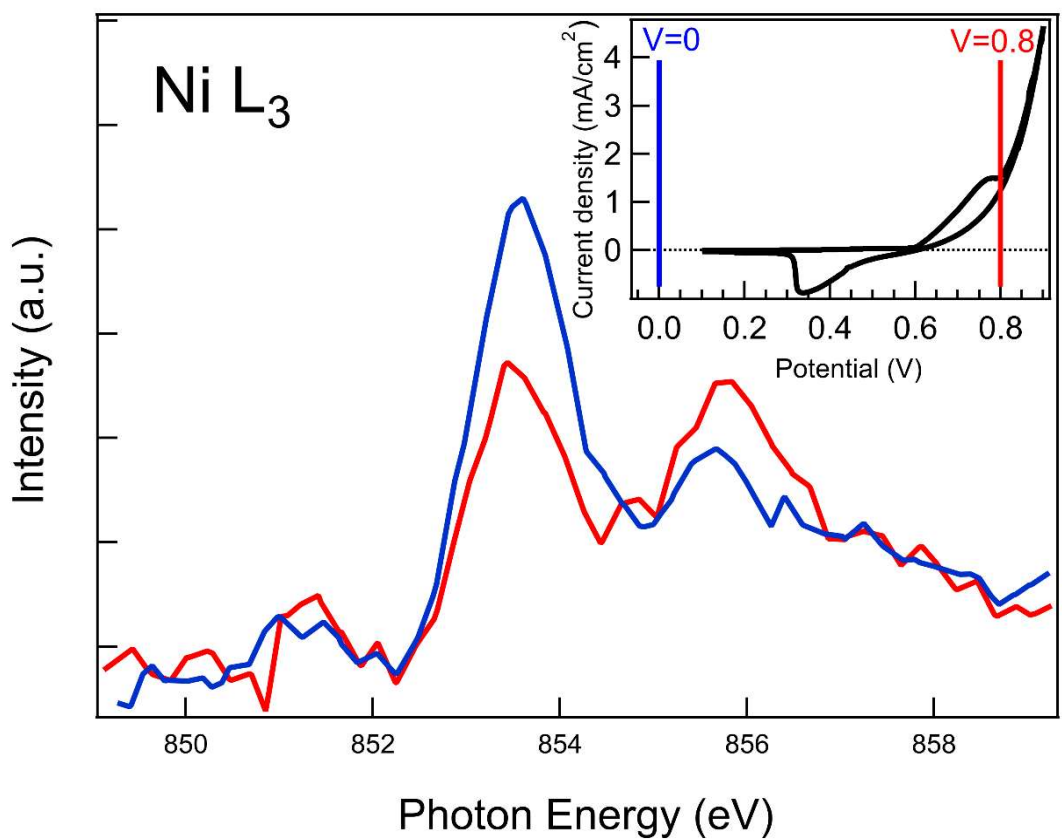

open source software developed by Stavitski and de Groot[46] (figure S6).

Fig. 8: *in operando* Ni $L_3$-edge XAS spectra measured in total fluorescence yield. The blue spectrum is recorded at an applied potential below the reduction potential (V=0), while the red curve is measured at an applied potential above the oxidation potential (V=0.8V). The inset reports the CV curve of the $Fe_xNi(OH)_2/Fe_xNiOOH$.

The CV of the sample measured at RT, in 1 M KOH, at a scanning rate of 0.02 V/s is shown in the inset of figure 8. The peak centered around 0.78 V can be assigned to the oxidation from Ni(II) to Ni(III),[47] and the onset of oxygen evolution reaction (OER) at potential > 0.8 V is also visible.[48]

We avoided to increase the potential to values higher than 0.9 V because the generation of oxygen bubbles at the surface of the catalyst may cause the rupture eof the $Si_3N_4$ membrane.

The CVs of Ni and Fe-Ni catalysts are comparable to those reported in literature[35,36,41,40].

Decreasing the potential, the peak at around 0.35V is assigned to the reduction of Ni (III) to Ni(II). Comparing the CVs of Ni (fig. 6), and Fe-Ni catalysts (fig. 8), it can be noticed that the potential difference between cathodic and anodic waves in Fe-Ni catalyst CV is larger than the one in Ni catalyst CV. This behavior can be attributed to a larger cell resistance in the Fe-Ni sample (the used potentiostat was not equipped with correction for uncompensated resistance). The addition of Fe to Ni catalystis known to decrease water oxidation overpotential, and to shift the Ni(II)/(III) redox waves to higher anodic potential.

The potential was increased above the oxidation potential from 0 V to 0.8 V at a scanning rate of 0.02 V/s and then kept constant during the XAS measurement (red solid line) of the Ni L3-edge. After that, in the same way the potential was decreased from 0.8 V to 0 V and then kept at this last potential during the acquisition of the spectrum (blue solid line). The line-shape of the two spectra is characterized by two peaks centered at 853.6 eV (A) and 855.5 eV (B) with different intensity ratio, reflecting the change in the Ni oxidation state.

In agreement with previous reported results[37,49] on a similar catalyst, at a potential higher than 0.7 V the increasing of the peak B is related mostly to the formation of $Ni^{3+}$ (see also XAS simulated spectra in figure S6).

**Conclusions**

We presented static and microfluidic cells for soft X-ray spectroscopy in the presence of fluids, designed and fabricated for the CNR-IOM BACH beamline at Elettra synchrotron radiation facility in Italy.

Examples of XAS measurements performed using the fluorescence yield on the O K-edge of different samples in aqueous solution and air were performed in the static cell. The design of the microfluidic cell equipped with three-electrodes allows to perform cyclic voltammetry, electrodeposition and electrochemical reactions *in operando* conditions. As a demonstration of its possibilities, we reported the cathodic electrodeposition of $Ni(OH)_2$ layers with variable thickness. The morphology of the deposited films was characterized by SEM evidencing the typical porous structure of the material. Ni L-edge XAS measurements revealed the 2+oxidation state in the electronic structure of $Ni(OH)_2$. A similar material with higher electrocatalyic performances due the presence of Fe($Fe_xNiOOH$) was used to test the cell for XAS measurements performed under *operando* conditions. The XAS spectra recorded on the Ni L-edge at applied potentials above the oxidation and below the reduction potentials show the expected change in Ni oxidation state.

The presented cells were designed on the basis of already existing cells and revealed good reliability and versatility. We expect that the availability of these cells at BACH beamline will raise a large interest in the user community for *in situ/operando* experiments on a variety of (electro)catalysts for different applications.

**Supplementary material**

Supplementary material reporting the full description of the cells is available.

**Acknowledgments:** S.N. andE.M acknowledge Dr. Junko Yano and Dr. Francesca Toma for hosting them as visiting scientists at Joint Center for Artificial Photosynthesis (JCAP) in Berkeley, Dr. Jinghua Guo and Dr. Ethan Crumlin for hosting them as visiting scientists on beamlines 6.3.1.2 and 9.3.1 at the Advanced Light Source (ALS) in Berkley. S.N, I.P, F.B, and E.M. acknowledge funding from IOM-CNR start up project 2010, EUROFEL project (RoadMap Esfri), MIUR project FIRB Futuro in Ricerca 2012 N. RBFR128BEC. S.N., E.M., and M.F. acknowledge funding from CNR-STM (Short Term Mobility) (projects 2015, 2016 (S.N. and E.M.) and 2016, 2017 (M.F.)) L.DA. acknowledges funding from Swedish Research Council (grant n. 2019-00663) and from the EU-H2020 Research and Innovation program under grant agreement n. 654360 Nanoscience Foundries and Fine Analysis-Europe.1

# Supporting information

# Soft X-ray spectroscopies in liquids and at solid-liquid interface at CNR-IOM BACH beamline at Elettra

S. Nappini, L. D'Amario, M. Favaro, S. Dal Zilio, F. Salvador, E. Betz- Güttner, A. Fondacaro, I. Píš, L. Romanzin, A. Gambitta, F. Bondino, M. Lazzarino, E. Magnano

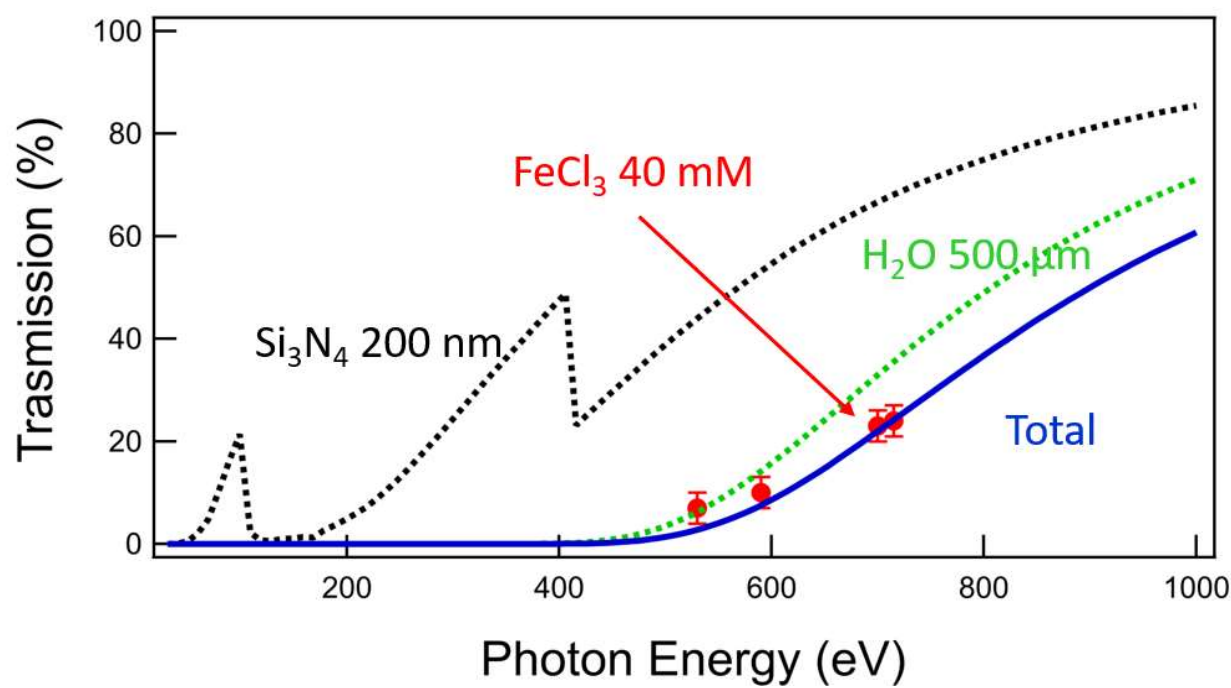

Fig. S1: Example of signal measured in the static cell for XAS in transmission: the red circles represent the intensity of the signal transmitted by the cell filled with an aqueous solution of FeCl$_3$ 40 mM in the 450–750 eV energy range. The signal from the solution is measured on the Au plate placed on the back of the cell and it is normalized to the incoming signal measured on the same Au plate hit by the direct beam.

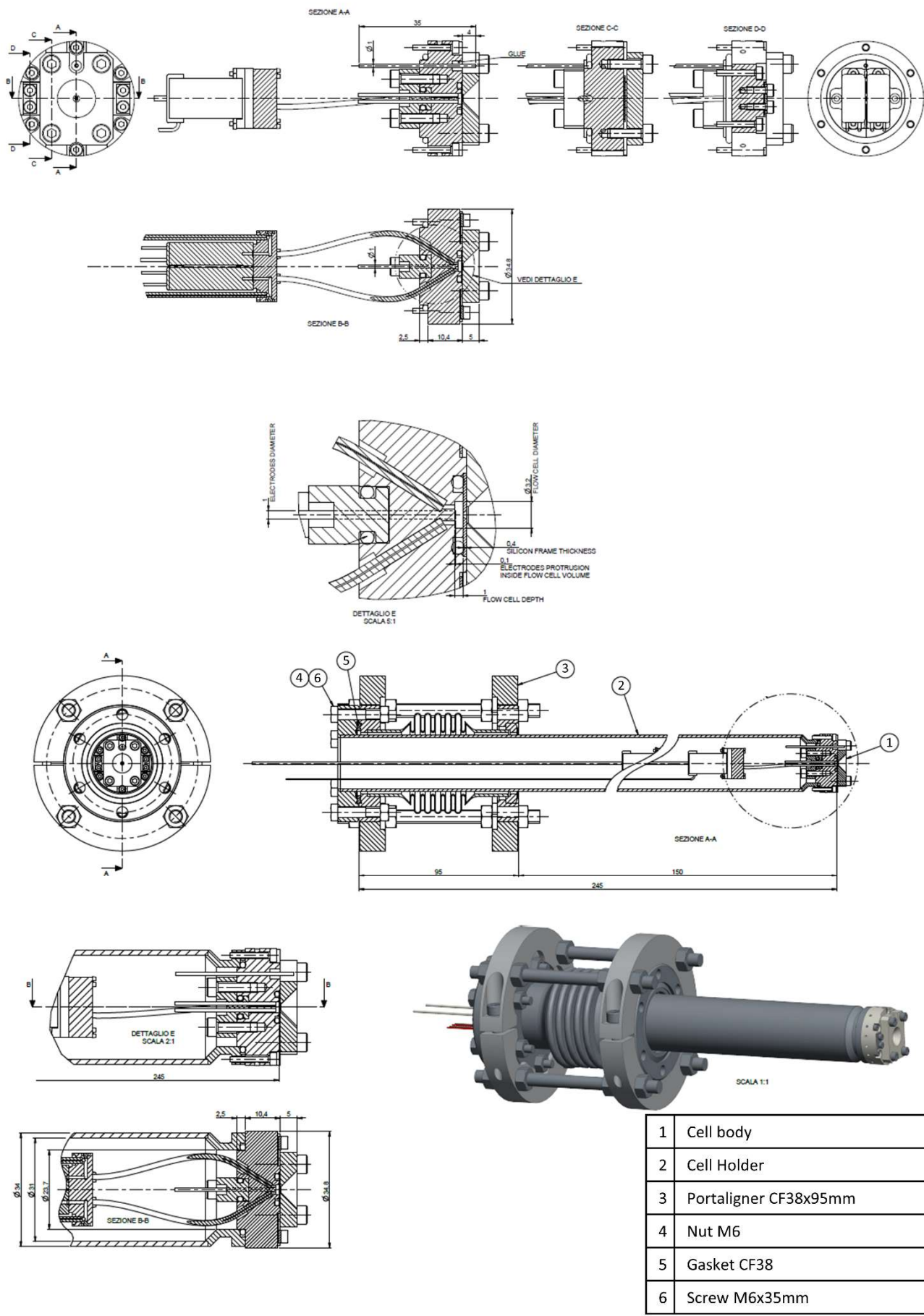

Fig. S2: Technical project of the microfluidic cell equipped with a three-electrode system.

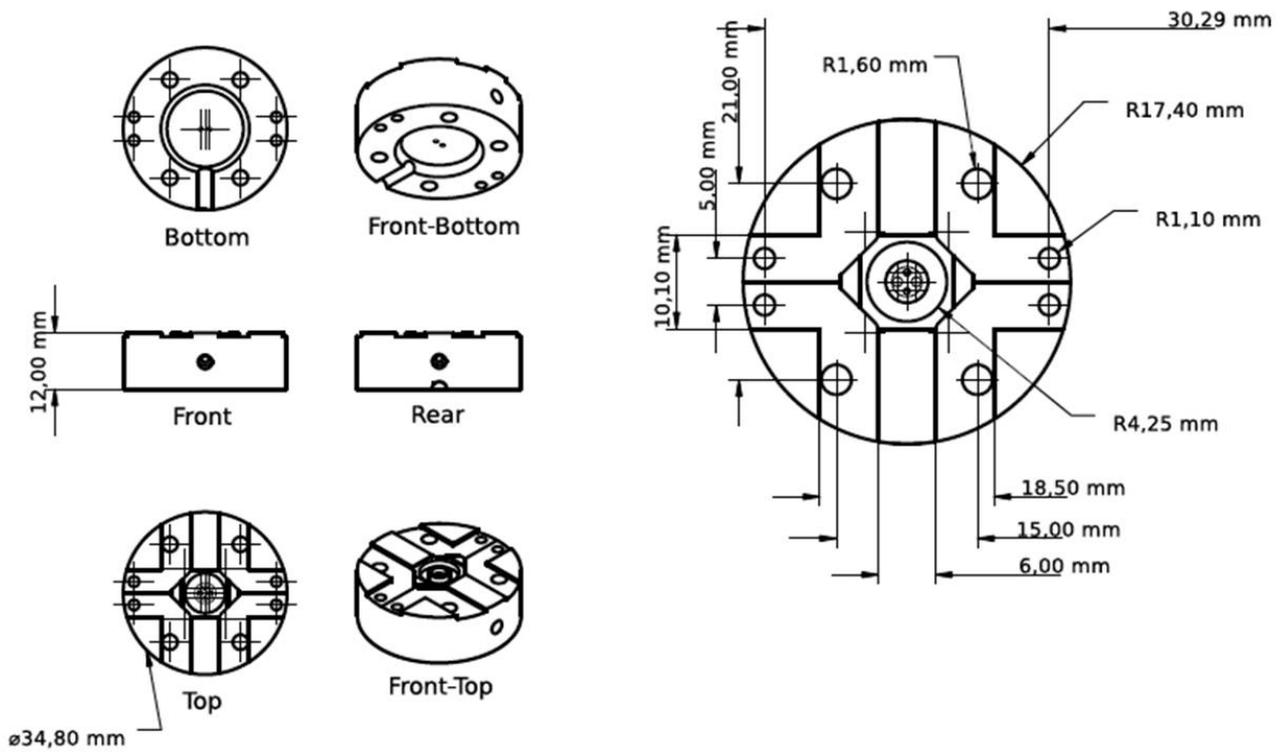

Fig. S3: Technical project of the microfluidic cell for 3D printer

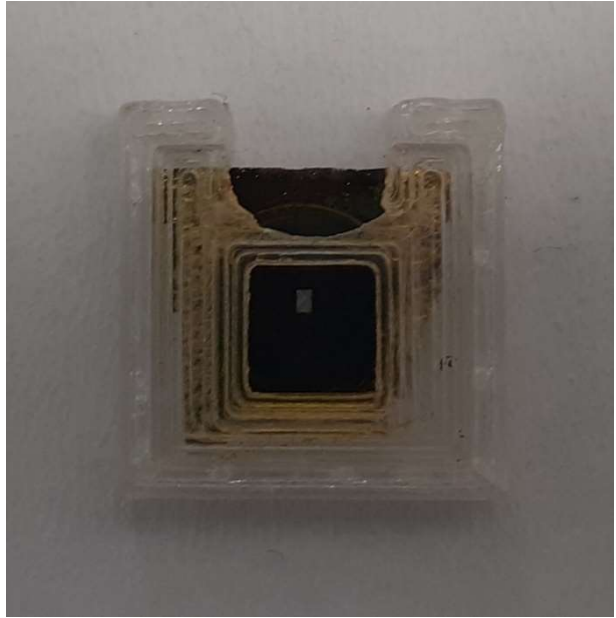

Fig. S4: Picture of the inert plastic mask containing the $Si_3N_4$/Ti/Au membrane

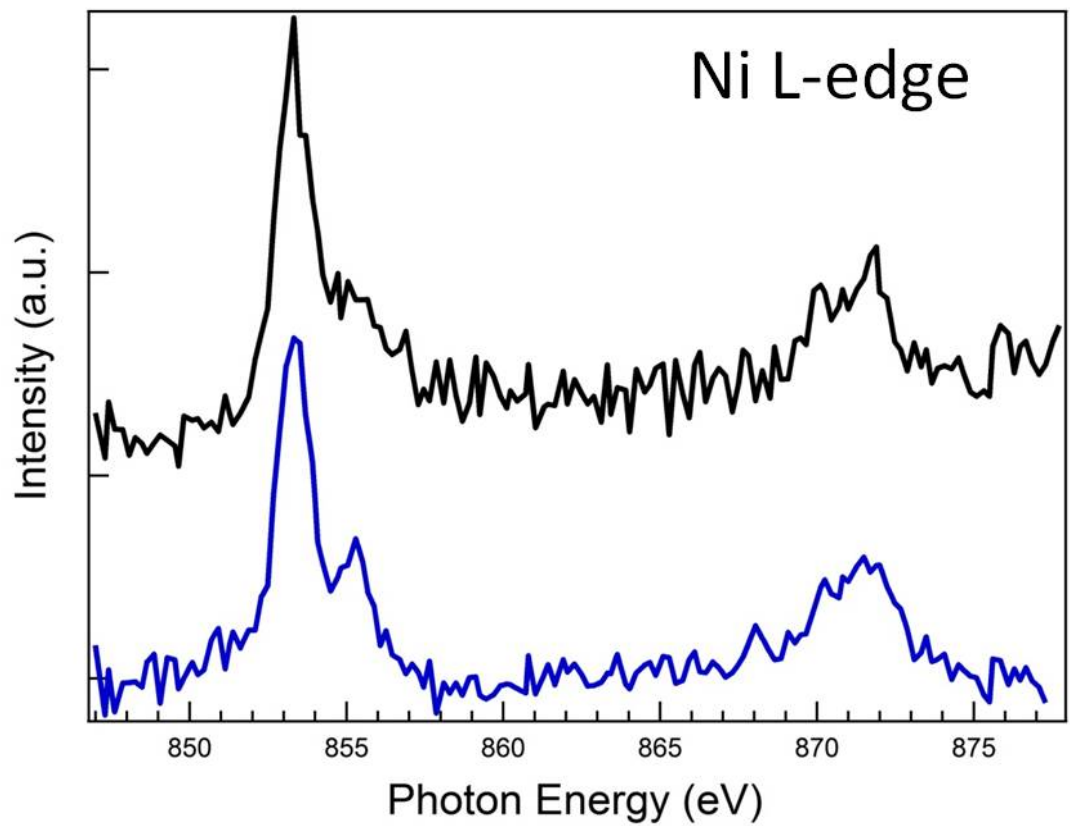

Fig. S5: Ni L-edge XAS spectra measured at open circuit potential in total fluorescence yield using MCP (bottom curve) and photodiode (top curve) detectors.

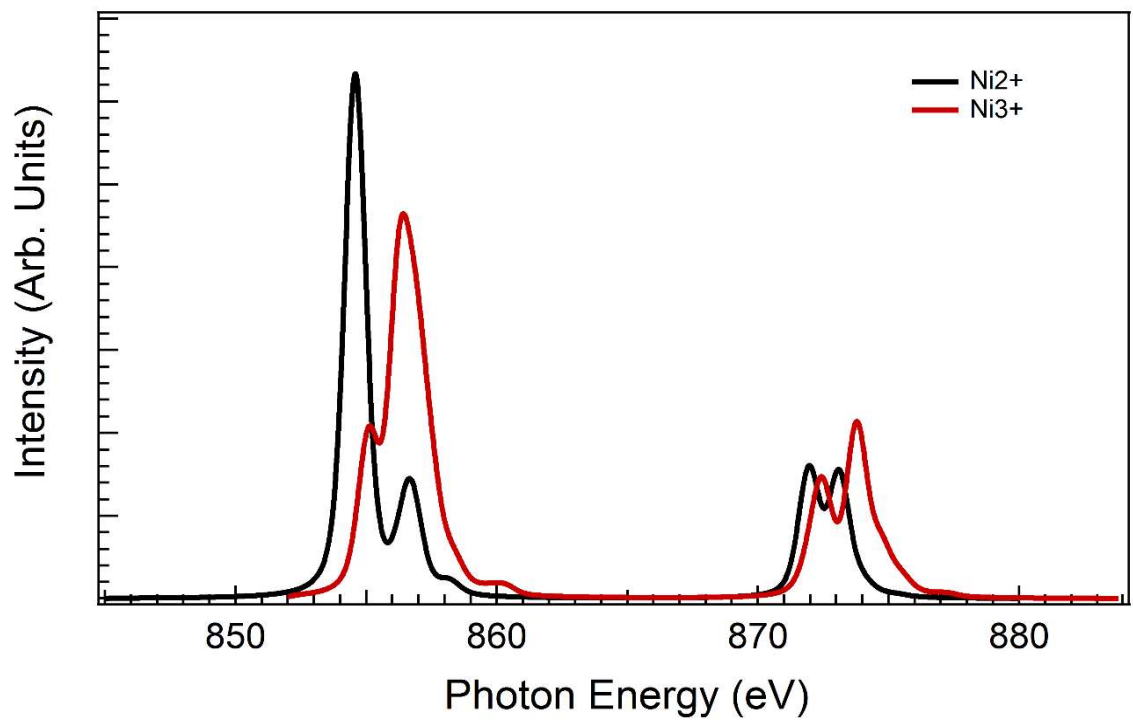

Fig. S6: Theoretical simulations of Ni L-edge spectra based on the Ligand Field Multiplet (LFM) theory carried out by CTM4XAS 5.5 open source software developed by Stavitski and de Groot [1]. Ni 2+ and 3+ was simulated in an octahedral symmetry (Oh) with 10Dq=2 and 3, respectively. Charge transfer effects were taken into account by a reduction of the Slater integrals in the calculation.

[1] E. Stavitski and F.M.F. de Groot, Micron **41**, 687 (2010).